\documentclass[3p,times,twocolumn]{elsarticle}
 \biboptions{comma,sort&compress}

\usepackage{graphicx}
\usepackage{amsmath,amssymb}
\usepackage{here}
\usepackage{ecrc}
\usepackage{physics}
\usepackage{adjustbox}

\newcommand{\eg}{\textit{e.g.,~}}
\newcommand{\ie}{\textit{i.e.,~}}

\newcommand{\gev}{\ensuremath{\,\mathrm{GeV}}}
\usepackage{xcolor}
\usepackage{ulem}

\usepackage{comment}

\volume{00}

\firstpage{1}

\journalname{Nuclear and Particle Physics Proceedings}
\runauth{Siyuan Li}

\jid{nppp}
\jnltitlelogo{Nuclear and Particle Physics Proceedings}
\usepackage{amssymb}
\usepackage[figuresright]{rotating}

\begin{document}

\begin{frontmatter}

\title{
Bounds on $a_\mu^{\mathrm{HVP,LO}}$ using H\"older's inequalities and finite-energy QCD sum rules\,$^*$} 
 
 \cortext[cor0]{Talk presented by Siyuan Li at QCD24 - 27th International Conference in Quantum Chromodynamics (8-12 July 2024,
  Montpellier - FR). }
  \author[label1]{Siyuan Li}
\address[label1]{Department of Physics and Engineering Physics, University of Saskatchewan, Saskatoon, SK, S7N~5E2, Canada}
\ead{siyuan.li@usask.ca}

\author[label1]{T.G. Steele}
\ead{tom.steele@usask.ca}

\author[label2]{J. Ho}
\address[label2]{Department of Physics, Dordt University, Sioux Center, Iowa, 51250, USA}
\ead{jason.ho@dordt.edu}

\author[label3]{R. Raza}
\address[label3]{Department of Physics, Thompson Rivers University, Kamloops, BC, V2C~0C8, Canada}
\ead{rraza@tru.ca}

\author[label3]{K. Williams}
\ead{williamsk16@mytru.ca}

\author[label3]{R.T. Kleiv}
\ead{rkleiv@tru.ca}

\pagestyle{myheadings}
\markright{ }
\begin{abstract}
\noindent
This study establishes
bounds on the leading-order (LO) hadronic vacuum polarization (HVP) contribution to the anomalous magnetic moment of the muon ($a_\mu^{\mathrm{HVP,LO}}$, $a_\mu = (g-2)_\mu/2$) by using H\"older's inequality and related inequalities in Finite-Energy QCD sum rules. Considering contributions from light quarks ($u,d,s$) up to five-loop order in perturbation theory within the chiral limit, leading-order light-quark mass corrections, next-to-leading order for dimension-four QCD condensates, and leading-order for dimension-six QCD condensates, the study finds QCD lower and upper bounds as $\left(657.0\pm 34.8\right)\times 10^{-10}\leq a_\mu^{\mathrm{HVP,LO}} \leq \left(788.4\pm 41.8\right)\times10^{-10}\,$.
 
%% keywords
\begin{keyword}  muon $g-2$, HVP, QCD sum rules, Finite-Energy QCD sum rules.

\end{keyword}

\end{abstract}
\end{frontmatter}

\section{Introduction} \label{sec:intro}
The muon $g-2$ experiment, one of the most significant current topics in particle physics, offers high-precision measurements that could potentially lead to the discovery of new physics beyond the Standard Model. 
A notable example of this experimental effort is Fermilab's recent announcement of the measurement of $a_\mu \equiv (g-2)_\mu/2$ in the summer of 2023~\cite{PhysRevLett.131.161802}. 
A crucial ingredient for the $a_\mu$ theoretical prediction is the leading-order (LO) hadronic vacuum polarization (HVP) contribution $a_\mu^\mathrm{HVP,LO}$. 
However, the theoretical value for $a_\mu^\mathrm{HVP,LO}$  remains an open question. 
Using data-driven dispersive methods to determine $a_\mu^\mathrm{HVP,LO}$, Ref.~\cite{AOYAMA20201} leads to  a $5.0\sigma$  discrepancy  with the experimental result from~\cite{PhysRevLett.131.161802}. 
The data-driven dispersive method from the CMD-3 collaboration~\cite{3collaboration2023measurement}, using pion form factor measurement, updates the low-energy $2\pi$ channel hadronic contribution to $a_\mu^\mathrm{HVP,LO}$, leading to a $0.9\sigma$ agreement with \cite{PhysRevLett.131.161802}. 
Lattice QCD (LQCD) approaches provide additional determinations of $a_\mu^\mathrm{HVP,LO}$ (\eg Ref.~\cite{Kuberski:2023qgx}).  In particular, the Budapest–Marseille–Wuppertal (BMW) result~\cite{Borsanyi:2020mff} for $a_\mu^\mathrm{HVP,LO}$, a high-precision outcome awaiting confirmation from other LQCD collaborations, introduces another tension with the data-driven determination of Ref.~\cite{AOYAMA20201}.

The discrepancy between theoretical methods for the hadronic contributions  to $a_\mu^\mathrm{HVP,LO}$
motivates our research to seek complementary bounds using QCD sum rules.
We provide an advanced theoretical prediction for $a_\mu^{\mathrm{HVP,LO}}$, the dominant source of uncertainty, using QCD sum rules as our framework. We establish upper and lower bounds for $a_\mu^{\mathrm{HVP,LO}}$ using H\"older's inequalities and finite-energy QCD sum rules (FESR). 
In Sec.~\ref{sec:method general}, we relate the dispersion integral for $a_\mu^\mathrm{HVP,LO}$~\cite{AOYAMA20201,PhysRev.174.1835,Steele1991} to the structure of FESR~\cite{Floratos:1978jb,Hubschmid:1980rm,Bertlmann:1984ih,Caprini:1985ex}, specifically introducing 
both upper and lower constraints for the
theoretical FESR quantity $F_{-2}(s_0)$ with parameter $s_0$, which allows for optimization of our theoretical prediction.
We can then extend H\"older's inequalities to FESR, 
thereby providing lower bound expression for $a_\mu^\mathrm{HVP,LO}$ in Sec.~\ref{sec:lower-bound} with suitable parameter inputs from Sec~\ref{sec:QCD_input}. 
We also developed an inequality-based methodology for deriving upper bounds (see Sec.~\ref{sec:upper-bound}), building on and expanding the approach in Ref.~\cite{Dalfovo:1992dr}, which exploits the synergy between H\"older inequalities and FESR.
Our prediction for $a_\mu^\mathrm{HVP,LO}$ includes the light-quark contributions five-loop
order in perturbation theory in the chiral limit, LO in light-quark mass corrections, next-to-leading
order (NLO) in dimension-four QCD condensates, and LO in dimension-six QCD condensates. The resulting upper and lower bounds bridge the gap between LQCD and data-driven approaches~\cite{AOYAMA20201}.
This paper summarizes the work of Ref.~\cite{Li:2024frm}; additional details and analysis can be found therein.

\section{Methodology}\label{sec:method general}

\begin{figure}[htb]
\centering
    \includegraphics[width = 0.95\linewidth]{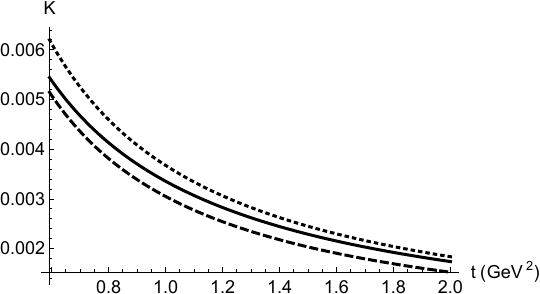}
    \caption{The exact $K(t)$ (solid line)  compared to the approximate form $K_\xi(t)$ with $\xi=0.83$ (lower dashed line) and with $\xi=1$ (upper dotted line).}
    \label{K_approx_fig}
\end{figure}

The dispersive approach gives $a_\mu^\mathrm{HVP,LO}$ as~\cite{AOYAMA20201,PhysRev.174.1835,Steele1991}
\begin{equation}
 a_\mu^{\mathrm{HVP,LO}}= \frac{\alpha^2}{3\pi^2}\int_{4m_\pi^2}^{\infty}\frac{1}{t}R(t)K(t)\,\mathrm{d}t\,,
 \label{eq:a_HVP_2}
\end{equation}
where $\alpha$ is the fine-structure constant.
Eq.~\eqref{eq:a_HVP_2} neglects terms of ${\cal O}\left(\frac{1}{t^3}\right)$.
The hadronic $R$-ratio can be written using the hadronic vacuum polarization spectral function as $R(t) = 12\pi \mathrm{Im}\Pi^H(t)$ \cite{Caprini:1984ud,Steele1991}.
The kernel function $K(t)$ can be approximated as
\begin{equation}
        \hspace{-0.7cm}
        K(t) = \int_{0}^{1} dx\, \frac{x^2(1-x)}{x^2+(1-x)t/m_\mu^2}\,
        \approx \frac{m_\mu^2}{3t}=K_{\rm approx}(t).
    \label{eq:K_approx}
\end{equation}
Thus, the QCD expression for Eq.~\eqref{eq:a_HVP_2} in terms of the hadronic spectral function becomes:
\begin{equation}
    a_\mu^\mathrm{QCD} \approx \frac{4 m_\mu^2\alpha^2}{3\pi} \int_{4m_\pi^2}^\infty \frac{1}{t^2} \mathrm{Im}\Pi^H\left(t\right) \mathrm{d}t\,.
    \label{eq:a_mu_QCD}
\end{equation}
Note that the $K_{\rm approx}(t)$ in Eq.~\eqref{eq:K_approx} provides an overestimate for $K(t)$ (see Fig.~\ref{K_approx_fig}). Consequently, this results in an upper bound for $a_\mu^\mathrm{QCD}$ derived from Eq.~\eqref{eq:a_HVP_2} and \eqref{eq:a_mu_QCD}:
\begin{align}\label{eq:upper_bound_general}
             a_\mu^\mathrm{QCD} &\le \frac{4 m_\mu^2\alpha^2}{3\pi} \int_{t_0}^\infty \frac{1}{t^2}  \mathrm{Im}\Pi^H\left(t\right) \mathrm{d}t
         \,,\quad t_0=4m_\pi^2\, .
\end{align}

To establish a lower bound for $a_\mu^\mathrm{QCD}$, it is crucial to obtain and verify a kernel function approximation that serves as an underestimate. This aspect is essential to ensure the accuracy of the lower bound. By focusing on the lower energy region from the threshold up to the $\rho$, $\omega$ peak, we introduce a scaling factor $\xi$ for $K(t)$ 
\begin{equation}
 K_\xi(t)=\xi K_{\mathrm{approx}}(t)=\xi  \frac{m_\mu^2}{3t}\,,
 \label{K_xi}
\end{equation}
where $\xi=0.83$. As shown in Fig.~\ref{K_approx_fig}, this value of $\xi$, derived using a standard Breit-Wigner $\sigma^\mathrm{BW}$ approach, ensures that the integral of $K_\xi(t)$ underestimates that of $K(t)$ across the energy range from the threshold $t_0$ to the $\rho$, $\omega$ peak, and provides $K_\xi(t)\le K(t)$ beyond the $\rho$, $\omega$ peak (see Ref.~\cite{Li:2024frm} for details).
By combining Eq.~\eqref{eq:a_mu_QCD} and \eqref{K_xi}, the lower bound is expressed as
\begin{align}
    a_\mu^\mathrm{QCD} &\ge \xi\frac{4 m_\mu^2\alpha^2}{3\pi} \int_{t_0}^\infty \frac{1}{t^2} \mathrm{Im}\Pi^H\left(t\right) \mathrm{d}t\,
    ,\quad t_0=4m_\pi^2\,.
    \label{eq:lower_bound_general}
\end{align}
The next step is to implement FESR, H\"older's inequalities, as well as the new inequality methodology adapted from Ref.~\cite{Dalfovo:1992dr} on Eq.~\eqref{eq:lower_bound_general} and \eqref{eq:upper_bound_general} to derive theoretical predictions for the lower and upper bounds of $a_\mu^{\mathrm{HVP,LO}}$.

\subsection{Lower Bounds}
\label{sec:lower-bound}

The hadronic representation of FESR is defined by \cite{Floratos:1978jb,Hubschmid:1980rm,Bertlmann:1984ih,Caprini:1985ex} 
\begin{equation}
F_{k}\left(s_{0}\right) = \int_{t_0}^{s_0} \frac{1}{\pi}\mathrm{Im}\Pi^H \left(t\right) t^k\, \mathrm{d}t\,
\label{eq:fesr}
\end{equation}
with integer $k$ indicating the weight of the sum rules and $t_0$ represents a physical threshold. Since Eq.~\eqref{eq:lower_bound_general} and \eqref{eq:fesr} share similar integral structures, we can express $a_\mu^\mathrm{QCD}$ using FESR by setting $k= -2$. Given that $\mathrm{Im}\Pi^H \left(t\right)$ is positive, this allows us to establish a lower bound for $F_{-2}$ with respect to parameter $s_0$:
\begin{equation}
    a_\mu^\mathrm{QCD} \ge \xi \frac{4 m_\mu^2\alpha^2}{3} F_{-2}\left(\infty\right) \ge \xi \frac{4 m_\mu^2\alpha^2}{3} F_{-2}\left( s_0\right) \,.
    \label{eq:a_mu_QCD_tom}
\end{equation}
We introduce the dependence of $s_0$ in our bound expression to allow optimization on the theoretical prediction, which will be detailed further in this section. However, FESR is only defined for integer weight $k \geq 0$ meaning  $F_{-2}$ does not have a direct theoretical prediction. In this research, we apply H\"older's, Cauchy-Schwarz, and related inequalities, while also developing additional inequality methodologies, to establish the fundamental properties of the field-theoretical result on QCD lower and upper bounds on $a_\mu^{\mathrm{HVP,LO}}$. 

H\"older's inequality has been adapted and developed within sum rules methodologies in previous studies (see \eg~Refs.~\cite{Benmerrouche:1995qa,Ho:2018cat,Steele:1998ry,Shi:1999hm,Wang:2016sdt,Yuan:2017foa}). It takes the general form of \cite{Beckenbach:1961, Berberian:1965}
\begin{align}
&\abs{\int_{t_1}^{t_2}f\left(t\right)g\left(t\right)\mathrm{d}\mu} \leq 
\left( \int_{t_1}^{t_2}\abs{f\left(t\right)}^{p}\mathrm{d}\mu \right)^{\frac{1}{p}}
\left( \int_{t_1}^{t_2}\abs{g\left(t\right)}^{q}\mathrm{d}\mu \right)^{\frac{1}{q}},\notag\\
&\qquad \qquad \frac{1}{p} + \frac{1}{q} = 1 \,.\label{eq:holder_general}
\end{align}
By substituting $f(t)$, $g(t)$ and $\mathrm{d}\mu=\frac{1}{\pi}\mathrm{Im}\Pi^H\left(t\right) \mathrm{d}t$,
\begin{align}
\abs{\int_{t_0}^{s_0}t^{\alpha+\beta} \frac{1}{\pi}\mathrm{Im}\Pi^H\left(t\right) \mathrm{d}t} \leq &
\left( \int_{t_0}^{s_0}\abs{t^{\alpha}}^{p}\frac{1}{\pi}\mathrm{Im}\Pi^H \left(t\right) \mathrm{d}t \right)^{\frac{1}{p}} \notag\\
&\times \left( \int_{t_0}^{s_0}\abs{t^{\,\beta} }^{q}\frac{1}{\pi}\mathrm{Im}\Pi^H\left(t\right) \mathrm{d}t \right)^{\frac{1}{q}}
\label{eq:fesr_tom}
\end{align}
which can be written in terms of FESR:
\begin{equation}
    F_{\alpha + \beta} \left(s_{0}\right)\leq \left[F_{\alpha \,p}\left(s_{0}\right)\right]^{\frac{1}{p}} \left[F_{ \frac{\beta\,p}{p-1}}\left(s_{0}\right)\right]^{\frac{p-1}{p}} \,.
    \label{eq:fesr-holder}
\end{equation}
We have explored several combinations of $\alpha$, $\beta$ and $p$ to construct possible lower bounds for $F_{-2}(s_0)$, which in turn provide lower bounds for $a_\mu^\mathrm{QCD}$ as per Eq.~\eqref{eq:a_mu_QCD_tom}. All considered combinations have integer weights in the range $0 \leq k \leq 2$ which helps to avoid dependence on unknown higher-dimension QCD condensates.The bounds are expressed as follows:
\begin{equation}
F_{-2} \geq \frac{F_{0}^3}{F_1^2} \geq \frac{F_{0}^2}{F_2} \geq \frac{F_1^4}{F_{2}^3} \,,
    \label{eq:fesr-hierarchy}
\end{equation}
where the dependence on $s_0$ in the FESR expressions is supressed.
The hierarchy among the candidates in Eq.~\eqref{eq:fesr-hierarchy} is determined using the Cauchy-Schwarz inequality below, which is a special case of the H\"older's inequality in Eq.~\eqref{eq:fesr-holder} (with $p = q = 2$):
\begin{equation}\label{eq.C-S_for_s0}
    F_k^2 \leq F_{k+1} F_{k-1}\,\rightarrow \; \frac{F_k}{F_{k+1}} \leq \frac{F_{k-1}}{F_k}\,.
\end{equation}
Therefore, the most restrictive lower bound from Eq.~\eqref{eq:fesr-hierarchy} yields a bound on $a_\mu^\mathrm{QCD}$ as given in Eq.~\eqref{eq:a_mu_QCD_tom}
(refer to Ref. [56] for additional information)
\begin{equation}
 a_\mu^\mathrm{QCD} \ge  \xi\frac{4 m_\mu^2\alpha^2}{3} \frac{F_{0}^3\left(s_0\right)}{F_1^2\left(s_0\right)}   \,,~\xi=0.83\,.
   \label{eq:a_mu_QCD-fesr_xi}
\end{equation}
It is now evident that adjusting $s_0$ in the lower bound expression allows us to find the strongest QCD bound.

\subsection{Upper Bounds}\label{sec:upper-bound}
Similar to the approach used in Sec.~\ref{sec:lower-bound} for applying FESR, we transform the integrals into $F_k(s_0)$ throughout our analysis of the upper bound given by Eq.~\eqref{eq:upper_bound_general}.

Using techniques adapted from Ref.~\cite{Dalfovo:1992dr}, we begin with the following expression, which follows from the positivity of the hadronic spectral function $\mathrm{Im}\Pi^H\left(t\right)$:
\begin{equation}
    \int\limits_{t_0}^{s_0}\frac{1}{t}\left[1+A t\right]^2 \mathrm{Im}\Pi^H\left(t\right) \,\mathrm{d}t\ge 0\,.
\end{equation}
This yields an intermediate upper bound for $F_{-1}(s_0)$, which will be used later in the search for the bound on $F_{-2}(s_0)$ by extremizing $A$:
\begin{gather}
    F_{-1}\le F_{-1}^{(B)}=\frac{F_0}{t_0}-\frac{\left(F_1/t_0-F_0\right)^2}{\left(F_2/t_0-F_1\right)}\,,
    \label{fm1_B}
    \\
    F_2/t_0-F_1>0\,.
    \label{fm1_B_condition}
\end{gather}
The validity of \eqref{fm1_B} relies on satisfying the subsidiary condition given in \eqref{fm1_B_condition}.

We now examine the upper bound expression given in \eqref{eq:upper_bound_general}, which satisfies the following two possible inequality relations.
\begin{equation}
\begin{aligned}
        \int\limits_{t_0}^{s_0}\frac{1}{t^2}\left[1+A t\right]^2 &\mathrm{Im}\Pi^H\left(t\right) \,\mathrm{d}t\\ 
        & \le 
    \begin{cases}
        \frac{1}{t_0} \int\limits_{t_0}^{s_0}\frac{1}{t}\left[1+A t\right]^2 \mathrm{Im}\Pi^H\left(t\right) \,\mathrm{d}t\,\\
        \frac{1}{t_0^2} \int\limits_{t_0}^{s_0}\left[1+A t\right]^2 \mathrm{Im}\Pi^H\left(t\right) \,\mathrm{d}t
    \end{cases}
\end{aligned}
\end{equation}
All three integrals can be written in terms of FESR. 
By extremizing $A$ and applying the Cauchy-Schwarz inequality given in \eqref{eq.C-S_for_s0}, we obtain two corresponding upper bounds for $F_{-2}$:
\begin{gather}
    F_{-2} \le \frac{F_{-1}^{(B)}}{t_0}-\frac{\left(F_0/t_0-F_{-1}^{(B)}\right)^2}{\left(F_1/t_0-F_0\right)}\,,
   \label{fm2_B_1} 
    \\
    \scalebox{0.9}{$F_1/t_0-F_0>0\,, \left(F_0/t_0-F_{-1}^{(B)}\right)^2<\left(F_0/t_0-F_0^2/F_1\right)^2\,,$
    }
    \label{fm2_B_condition_1}
\end{gather}
and
\begin{gather}
     F_{-2} \le F_0/t_0^2-\frac{\left(F_1/t_0^2-F_{-1}^{(B)}\right)^2}{\left(F_2/t_0^2-F_0\right)}\,,
   \label{fm2_B_2}  
   \\
  \scalebox{0.9}{$F_2/t_0^2-F_0>0\,, \left(F_1/t_0^2-F_{-1}^{(B)}\right)^2<\left(F_1/t_0^2-F_0^2/F_1\right)^2,$}
  \label{fm2_B_condition_2}
\end{gather}
where \eqref{fm2_B_condition_1} serves as the subsidiary condition for \eqref{fm2_B_1}, and \eqref{fm2_B_condition_2} applies to \eqref{fm2_B_2}. 
Both forms of the upper bound, \eqref{fm2_B_1} and \eqref{fm2_B_2}, yield identical numerical results. Therefore, the upper QCD bound on $a_\mu^\mathrm{QCD}$ from \eqref{eq:upper_bound_general} is given by
(see Ref.~\cite{Li:2024frm} for more details)
\begin{equation}
   a_\mu^\mathrm{QCD}
   \le \frac{4 m_\mu^2\alpha^2}{3} 
   \begin{cases}
   F_{-1}^{(B)}/{t_0}-\frac{\left(F_0/t_0-F_{-1}^{(B)}\right)^2 }{F_1/t_0-F_0}
   \\[12pt]
    F_0/t_0^2-\frac{\left(F_1/t_0^2-F_{-1}^{(B)}\right)^2}{F_2/t_0^2-F_0}
   \,
   \end{cases}  
   .
   \label{eq:a_mu_QCD-fesr_upper}
\end{equation}
Combining this with our lower bound from Eq.~\eqref{eq:a_mu_QCD-fesr_xi}, the complete fundamental QCD constraint on the $a_\mu^\mathrm{LO,HVP}$ is (see Ref.~\cite{Li:2024frm} for more details)
\begin{align}
    \xi\frac{4 m_\mu^2\alpha^2}{3} \frac{F_{0}^3}{F_1^2} &\le   
 a_\mu^\mathrm{QCD}
   \le \frac{4 m_\mu^2\alpha^2}{3} 
   \begin{cases}
   F_{-1}^{(B)}/{t_0}-\frac{\left(F_0/t_0-F_{-1}^{(B)}\right)^2 }{F_1/t_0-F_0}
   \\[10pt]
    F_0/t_0^2-\frac{\left(F_1/t_0^2-F_{-1}^{(B)}\right)^2}{F_2/t_0^2-F_0}
   \end{cases}\!\!\!\!\!, \notag \\
    &\xi=0.83\,,
      \label{eq:a_mu_QCD-fesr_summary}  
\end{align}
where the $s_0$ dependencies are suppressed for all $F_k$'s. 
The parameter $s_0$ can be independently adjusted on either side of the inequality relations to determine the most restrictive bounds for $a_\mu^\mathrm{QCD}$. Both bounds are constructed with well-determined low-weight FESRs $\left(F_0\,, F_1\,, F_2 \right)$.
One might be inclined to interpret $s_0$ as either a cutoff for the QCD contributions or as a duality threshold, but these interpretations are incorrect. In our approach, $s_0$ is used to establish fundamental QCD bounds. The key constraints on $s_0$ 
are dictated by the fundamental inequalities in Eq.~\eqref{eq.C-S_for_s0} and the subsidiary conditions  outlined in Eqs.~\eqref{fm1_B_condition}, \eqref{fm2_B_condition_1} and \eqref{fm2_B_condition_2}. 

\section{Finite-Energy QCD Sum Rules Parameter Inputs}
\label{sec:QCD_input}
The correlation functions for the light quark ($u,d,s$) vector current, defined as $j_\mu(x)  = \bar{q}(x)\,\gamma_\mu \, q(x)$,
correspond to the hadronic spectral function in Eq.~\eqref{eq:fesr}
\begin{align}
    \Pi\left(Q^2\right) =   \frac{1}{4\pi^2} &\Pi^\mathrm{pert}\left(Q^2\right)
    - \frac{3 m_q^2(\nu)}{2\pi^2 Q^2} \nonumber\\
    &+ 2\langle m_q \bar{q}q\rangle\frac{1}{Q^4}\left(1+\frac{1}{3}\frac{\alpha_s(\nu)}{\pi}\right)
    \nonumber \\
    & +\frac{1}{12\pi}\langle \alpha_s G^2 \rangle \frac{1}{Q^4}
    \left(1+\frac{7}{6}\frac{\alpha_s(\nu)}{\pi}\right)\nonumber\\
    & -\frac{224}{81}\pi \alpha_s \langle \bar{q}\bar{q}qq\rangle \frac{1}{Q^6}\,,
    \label{eq:correlator}
\end{align}
where the QCD correlation function is extended to five-loop order in $\overline{\mathrm{MS}}$-scheme perturbation theory in the chiral limit \cite{PhysRevLett.101.012002,Gorishnii:1990vf,Surguladze:1990tg,Chetyrkin:1996ez,Chetyrkin:1979bj,Dine:1979qh,Celmaster:1979xr}, 
LO in light-quark mass corrections \cite{Reinders:1984sr,Pascual1984,Narison:2002woh,Steele1991},
NLO in dimension-four QCD condensates \cite{Surguladze:1990sp,Wang:2016sdt,Chetyrkin:1985kn}, 
and to LO in dimension-six QCD condensates \cite{Shifman:1978bx,Shifman:1978by,Reinders:1984gu} (note that the mass corrections  known to three-loop order \cite{Gorishnii:1986pz} have not been included and will be considered in future work).
The correlation function is for single flavour and requires a quark-charge pre-factor $Q_q^2$, where $Q_u^2 = 4/9$ and $Q_d^2 = Q_s^2 = 1/9$.
The unprecedented higher-order correction in the perturbative contributions are from
\begin{gather}
  \frac{1}{\pi}\mathrm{Im}\Pi^\mathrm{pert}\left(t,\nu\right) =S\left[x(\nu),L(\nu)\right] = 1+\sum_{n=1}^{\infty} x^{n} \sum_{m=0}^{n-1} T_{n,m}L^m\,,
    \notag
    \\
    \mathrm{with} \quad x(\nu) \equiv \frac{\alpha_s(\nu)}{\pi}\,,~L(\nu) \equiv \log\left(\frac{\nu^2}{t}\right)\,.\label{eq:pertseries}
\end{gather}
The coefficients $T_{n,m}$ were derived up to $n=4$ with $N_f=3$ for five-loop order contribution and the values can be found in Table \ref{table:coeff_nf_3}. 
The renormalization scale from our energy range spans both the $N_f=4$ and $N_f=3$ regimes. However, as we discuss later in Sec.~\ref{sec:result}, the optimal choice of $s_0$ falls within $N_f=3$ region. Consequently, the values provided in Table \ref{table:coeff_nf_3} are specific to the $N_f=3$ regime and are used exclusively for the final results presented in Table \ref{table:result}
(see Ref.~\cite{Li:2024frm} for the $N_f=4$ values associated with Table~\ref{table:coeff_nf_3})
.

\begin{table}[ht]
\centering
\renewcommand{\arraystretch}{1.2}
\begin{tabular}{c c c c c }
\hline \hline
  $N_f = 3$ & $m = 0$ & $m = 1$ & $m = 2$ & $m = 3$\\ \hline
 $n = 1$ & 1 & -- & -- & --\\ 
 $n = 2$ & 1.63982 & 9/4 & -- & -- \\ 
 $n = 3$ & -10.2839 & 11.3792 & 81/16 & -- \\ 
 $n = 4$ & -106.896 & -46.2379 & 47.4048 & 729/64 \\ \hline \hline
\end{tabular}
\caption{
The imaginary part of the vector-current correlation function \eqref{eq:pertseries} perturbation contribution coefficient $T_{n,m}$ up to five-loop order in $\overline{\mathrm{MS}}$-scheme for $N_f = 3$. 
The four-loop results are detailed in Ref.~\cite{PhysRevD.67.034017}, the five-loop coefficient $T_{4,0}$ is given in \cite{PhysRevLett.101.012002}, and five-loop logarithmic coefficients $T_{4,1}$, $T_{4,2}$, and $T_{4,3}$ are derived from the renormalization group analysis in Ref.~\cite{PhysRevD.67.034017} via the four-loop $\overline{\mathrm{MS}}$-scheme $\beta$ function \cite{vanRitbergen:1997va}.}
\label{table:coeff_nf_3}
\end{table}

We require FESR for weight $k=\{0,1,2\}$ to calculate our constraint from Eq.~\eqref{eq:a_mu_QCD-fesr_summary}. 
These can be expressed using the FESR methodology \cite{Floratos:1978jb,Hubschmid:1980rm,Bertlmann:1984ih,Caprini:1985ex}, with a renormalization scale of $\nu = \sqrt{s_0}$:
\begin{align}
    F_0\left(s_0\right) &=  \frac{1}{4\pi^2}\Bigg[1  + \frac{\alpha_s(\nu)}{\pi}T_{1,0} + \left(\frac{\alpha_s(\nu)}{\pi}\right)^2\left(T_{2,0}+T_{2,1}\right) \nonumber\\
    & + \left(\frac{\alpha_s(\nu)}{\pi}\right)^3\left(T_{3,0}+T_{3,1}+2T_{3,2}\right) \nonumber\\ 
    & \hspace{-1cm}
    + \left(\frac{\alpha_s(\nu)}{\pi}\right)^4 \left(T_{4,0}+T_{4,1}+2T_{4,2}+6T_{4,3}\right) \Bigg]s_0
    -\frac{3}{2\pi^2}  m_q(\nu)^2\,,
    \label{eq:fesr-k-0}
\end{align}
\begin{align}
 F_1\left(s_0\right)& =  \frac{1}{8\pi^2}\Bigg[1 + \frac{\alpha_s(\nu)}{\pi}T_{1,0} + \left(\frac{\alpha_s(\nu)}{\pi}\right)^2\left(T_{2,0}+\frac{1}{2}T_{2,1}\right)\nonumber\\
 & + \left(\frac{\alpha_s(\nu)}{\pi}\right)^3\left(T_{3,0}+\frac{1}{2}T_{3,1}+\frac{1}{2}T_{3,2}\right) \nonumber\\ 
    & + \left(\frac{\alpha_s(\nu)}{\pi}\right)^4\left(T_{4,0}+\frac{1}{2}T_{4,1}+\frac{1}{2}T_{4,2}+\frac{3}{4}T_{4,3}\right)\Bigg]s_0^2 
    \nonumber\\
    & \hspace{-1cm}
    - 2\langle m_q \bar{q}q\rangle\left(1+\frac{1}{3}\frac{\alpha_s(\nu)}{\pi}\right)
    - \frac{1}{12\pi}\langle \alpha_s G^2 \rangle\left(1+\frac{7}{6}\frac{\alpha_s(\nu)}{\pi}\right)\,,
    \label{eq:fesr-k-1}
\end{align}
\begin{align}
    F_2\left(s_0\right) &=  \frac{1}{12\pi^2}\Bigg[1 + \frac{\alpha_s(\nu)}{\pi}T_{1,0} + \left(\frac{\alpha_s(\nu)}{\pi}\right)^2\left(T_{2,0}+\frac{1}{3}T_{2,1}\right)
    \nonumber\\ &
    + \left(\frac{\alpha_s(\nu)}{\pi}\right)^3\left(T_{3,0}+\frac{1}{3}T_{3,1}+\frac{2}{9}T_{3,2}\right) \nonumber\\ 
    & 
    + \left(\frac{\alpha_s(\nu)}{\pi}\right)^4\left(T_{4,0}+\frac{1}{3}T_{4,1}+\frac{2}{9}T_{4,2}+\frac{2}{9}T_{4,3}\right)\Bigg]s_0^3
    \nonumber \\&
    - \frac{224}{81} \pi \alpha_s \langle \bar{q}\bar{q}qq\rangle  \,.
    \label{eq:fesr-k-2}
\end{align}
All QCD parameters required to calculate Eqs.~\eqref{eq:correlator} and \eqref{eq:fesr-k-0}-\eqref{eq:fesr-k-2} are listed in Table ~\ref{table:parameters}.
\begin{table}[htb]
\centering
\resizebox{\columnwidth}{!}{
\renewcommand{\arraystretch}{1.3}
\begin{tabular}{l c c}
\hline \hline
\multicolumn{1}{l}{Parameter} & \multicolumn{1}{c}{Value} & \multicolumn{1}{c}{Source} \\ \hline
 $\alpha$ & $1/137.036$ & \cite{PDG2022} \\
 $\alpha_s\left(M_\tau\right)$ & $0.312\pm 0.015$ & \cite{PDG2022} \\
 $m_u(2\,\mathrm{GeV})$ & $2.16^{+0.49}_{-0.26}\,\mathrm{MeV}$ & \cite{PDG2022}\\
 $m_d(2\,\mathrm{GeV})$ & $4.67^{+0.48}_{-0.17}\,\mathrm{MeV}$ & \cite{PDG2022}\\
 $m_s(2\,\mathrm{GeV})$ & $\left(0.0934^{+0.0086}_{-0.0034}\right)\gev$ & \cite{PDG2022} \\
 $f_\pi$ & $\left(0.13056\pm 0.00019\right)/\sqrt{2}$ GeV& \cite{PDG2022} \\
 $m_n \langle \bar{n}n\rangle$ & $-\frac{1}{2}f_\pi^2 m_\pi^2$ & \cite{Gell-Mann:1968hlm}  \\
 $m_s \langle \bar{s}s\rangle$  & $r_m r_c m_n \langle\overline{n}n\rangle $  & \cite{Harnett2021} \\
 $r_c \equiv \langle \bar{s}s\rangle/\langle \bar{n}n\rangle$ & $0.66 \pm 0.10 $  & \cite{Harnett2021} \\
 $m_s/m_n = r_m$ & $27.33^{+0.67}_{-0.77}$ & \cite{PDG2022}  \\
 $\langle \alpha G^2 \rangle$ & $\left(0.0649\pm 0.0035\right)\gev^4$ & \cite{Albuquerque:2023bex}\\
  $\kappa$ & $3.22\pm 0.5$ & \cite{Albuquerque:2023bex} \\
 $ \alpha_s \langle\bar{n}n\rangle^2$ & $\kappa\left(1.8\times 10^{-4}\right)\gev^6$  & \cite{Harnett2021} \\
  $ \alpha_s \langle \bar{s}s\rangle^2$ & $r_c^2 \alpha_s \langle \bar{n}n\rangle^2$ & \cite{Harnett2021}\\
  \hline \hline
\end{tabular}
}
\caption{QCD parameters and uncertainties used in our calculations. Here, $m_n = \left(m_u+m_d\right)/2$ and $\langle\bar{n}n\rangle = \langle\bar{u}u\rangle = \langle\bar{d}d\rangle$.
 }
\label{table:parameters}

\end{table}

Again, the required FESRs for establishing the bounds are derived from the vector current correlation function \eqref{eq:correlator} for single light-quark flavour. Thus, Eqs.~\eqref{eq:fesr-k-0}--\eqref{eq:fesr-k-2} will be calculated for all light-quark flavours, incorporating the corresponding charge pre-factors $Q_q^2$.

\section{Calculation, Optimization and Results}
\label{sec:result}
With parameters provided in Table ~\ref{table:coeff_nf_3} and \ref{table:parameters}, we can calculate the FESRs defined in Eqs.~\eqref{eq:fesr-k-0}--\eqref{eq:fesr-k-2} to derive the lower and upper bounds on $a_\mu^\mathrm{QCD}$, as summarized by Eq.~\eqref{eq:a_mu_QCD-fesr_summary} [see also \eqref{eq:a_mu_QCD-fesr_xi} and \eqref{eq:a_mu_QCD-fesr_upper}]. 
The maximum value for Eq.~\eqref{eq:a_mu_QCD-fesr_xi} and the minimum value for Eq.~\eqref{eq:a_mu_QCD-fesr_upper} are obtained while the $k=1$ case of the Cauchy-Schwarz inequality \eqref{eq.C-S_for_s0} is satisfied. This ensures the validity of the FESRs as well as the integrated hadronic spectral function.
To determine the optimized $s_0^\mathrm{opt}$ value that provides the most restrictive bounds, we start from a high energy scale near the bottom threshold within the $N_f=4$ regime and proceed towards lower energy regions. We observe that lower $s_0^\mathrm{opt}$ values yield stronger bounds, leading us to shift to the $N_f=3$ regime below the charm threshold. A renormalization group (RG) analysis is also conducted to ensure that the FESRs remain self-consistent across flavor thresholds (see Ref.~\cite{Li:2024frm} for more details in RG analysis).

Additionally, we use a flavor-separated approach by determining the optimal $s_0^\mathrm{opt}$ for each flavor individually first. The contributions from each flavor are then combined with their respective charge pre-factors $Q_q^2$ to obtain the final bounds on $a_\mu^\mathrm{QCD}$.
This flavor-separated approach yields stronger bounds compared to using a single general $s_0^\mathrm{opt}$ for the combined expression of FESRs with charge pre-factors.
Thus, the final results are derived from the flavor-separated method. 
The values of $s_0^\mathrm{opt}$ for each flavor and the final bounds obtained through this approach are shown in Table ~\ref{table:result}. 

As discussed above it is important to note that $s_0^\mathrm{opt}$ represents the value that produces the most restrictive bounds while ensuring the validity of the FESRs via the fundamental inequality constraints \eqref{eq.C-S_for_s0} and the subsidiary conditions (\ref{fm1_B_condition},\ref{fm2_B_condition_1},\ref{fm2_B_condition_2}). Therefore, $s_0^\mathrm{opt}$ should not be interpreted as a cut-off for the QCD contributions or as a duality threshold. However, it is still important to be attentive to the perturbative convergence associated with $s_0^\mathrm{opt}$.  For example, in Eq.~\eqref{eq:fesr-k-0}, the relative perturbative contributions from LO (one loop) up to five loop are respectively $\{1,0.14,0.08,0.03, 0.004\}$ and hence $s_0^\mathrm{opt}$ leads to good convergence in the perturbative series. 

\begin{table}[htb]
\centering
\resizebox{\columnwidth}{!}{
\renewcommand{\arraystretch}{1.5}
\begin{tabular}{l c c c }
\hline\hline
Flavour & $s_0^\mathrm{opt}\,\left(\mathrm{GeV}^2 \right)$ & $a_\mu^\mathrm{QCD}$ (lower bound)
& $a_\mu^\mathrm{QCD}$  (upper bound)\\ \hline
$u$ & $1.09$ & $\ge 472.7\times 10^{-10}$  & $\le 567.2\times 10^{-10} $
\\
$d$ & $1.09$ & $\ge 118.1\times 10^{-10}$  & $\le 141.7\times 10^{-10} $
\\
$s$ & $1.19$ & $\ge 66.2\times 10^{-10}$ & $\le 79.5\times 10^{-10} $
\\
Total & -- & $\ge 657.0\times 10^{-10}$ & $\le 788.4\times 10^{-10} $
\\\hline\hline
\end{tabular}
}
\caption{The optimized $s^\mathrm{opt}_0$ and corresponding bounds on $a_\mu^\mathrm{QCD}$ are shown for each light-quark flavour using the flavour-separated method with central values of the QCD input parameters from Table~\ref{table:parameters}.  
The field-theoretical difference between the $u$ and $d$ quark contributions, which arises from the quark mass, is small, resulting in the same $s_0^\mathrm{opt}$ for both non-strange channels.
The ``Total" entry row presents the sum of the individual flavour contributions for the final predicted bounds on $a_\mu^\mathrm{QCD}$.}
\label{table:result}
\end{table}

Taking into account the uncertainties which is mainly contributed from the vacuum saturation parameter  $\kappa$ and the dimension-four gluon condensate parameter $\langle \alpha G^2\rangle$ (see Table \ref{table:parameters}), we present our final QCD prediction for the lower and upper bounds of the light-quark contributions to be (details available in Ref.~\cite{Li:2024frm}):

\scalebox{.95}{\parbox{\linewidth}{
\begin{align}
\hspace{-0.4cm}
\left(657.0\pm 34.8\right)\times10^{-10}\leq a_\mu^\mathrm{QCD} \leq \left(788.4\pm 41.8\right)\times10^{-10}.
\label{eq:a_mu_QCD-result}
\end{align}
}}

For comparison with data-driven approaches and LQCD results, we have included underestimated high-energy perturbative contributions above the charm threshold in our light-quark contributions. This supplementation accounts for charmonium and bottomonium resonances, with a total contribution of 
$a_{\mu\,,\,\bar{c}c\,,\,\bar{b}b}^{\mathrm{HVP,LO}}=(7.93\pm 0.19)\times 10^{-10}$, as reported in \cite{PhysRevD.101.014029}. 
Our final bounds, incorporating these contributions for comparison purposes, are given by (more details in  Ref.~\cite{Li:2024frm})

\begin{figure}[htb]
    \centering
    \includegraphics[width=\linewidth]{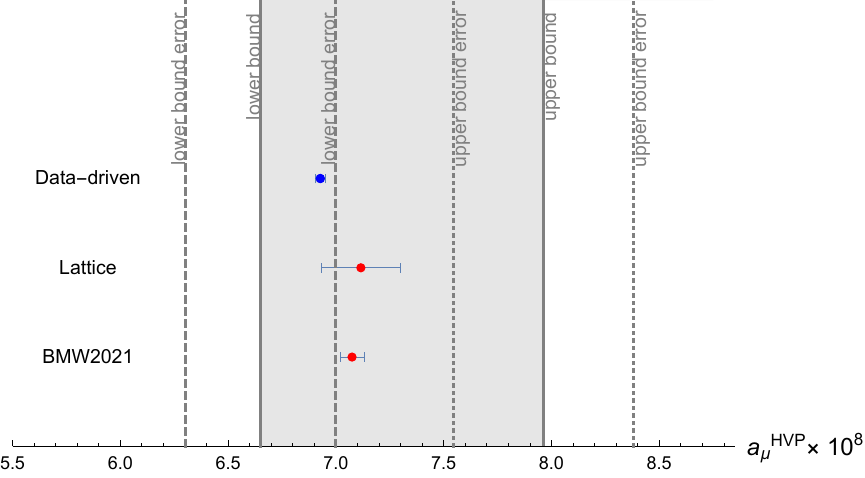}
    \caption{The $a_\mu^\mathrm{QCD}$ results in Eq.~\eqref{eq:inclusive_bound} include lower (long dashed lines) and upper bounds (short dashed lines) reflecting theoretical uncertainties.
    These are compared to the world theoretical averages for $a_\mu^{\mathrm{HVP,LO}}$ provided in \cite{AOYAMA20201}. The data-driven result is shown in blue, while the LQCD world average \cite{AOYAMA20201} and the BMW collaboration's sub-percent precision calculation \cite{Borsanyi:2020mff} are shown in red.
    The grey region indicates the allowed central-value range of our QCD predictions in Eq.~\eqref{eq:inclusive_bound}. 
    }
    \label{fig:comparison}
\end{figure}

\scalebox{.95}{\parbox{\linewidth}{
\begin{align}
\hspace{-0.4cm}
 \left(664.9\pm 34.8\right)\times10^{-10}  \leq a_\mu^\mathrm{HVP,LO}  \leq 
 \left(796.3\pm 41.8\right)\times10^{-10}.
 \label{eq:inclusive_bound}
\end{align}
}}
These results are compared in Fig.~\ref{fig:comparison} with the data-driven result from the ($g-2$) Theory Initiative Whitepaper  \cite{AOYAMA20201} 
\begin{equation}
    a_\mu^{\mathrm{HVP,LO}} = \left(693.1\pm 4.0\right)\times10^{-10}, 
     \label{eq:data-driven-result_2}
\end{equation}
and LQCD result from the ($g-2$) Theory Initiative Whitepaper \cite{AOYAMA20201}
\begin{equation}
    a_\mu^{\mathrm{HVP,LO}} = \left(711.6\pm 18.4\right)\times10^{-10}.
    \label{eq:lqcd-result_jason}
\end{equation}

As a supplementary observation, the inequality analysis conducted on FESRs in Sec.~\ref{sec:method general} is also applicable to Laplace sum rules (LSR)  (see Ref.~\cite{Li:2024frm}'s appendix for more detailed calculation and analysis). However, we found that FESR bounds are considerably stronger than those obtained from the LSR approach. Therefore, we will adhere to the FESR results. 

\section{Conclusion}
In light of the recent discrepancies among theoretical predictions for the LO HVP (hadronic) contributions to the muon anomalous magnetic moment using data-driven inputs from the CMD-3 collaboration~\cite{3collaboration2023measurement}, data-driven methods of \cite{Borsanyi:2020mff} and the latest LQCD results~\cite{Kuberski:2023qgx}, 
 we present our fundamental QCD theoretical result (\ie Eqs.~\eqref{eq:a_mu_QCD-result} and \eqref{eq:inclusive_bound})~\cite{Li:2024frm}. As shown in Fig.~\ref{fig:comparison}, our findings are consistent with both LQCD and data-driven estimates of $a_\mu^{\mathrm{HVP,LO}}$, 
presenting a potential avenue for resolving these discrepancies. 

In this research, we introduce a novel methodology that combines QCD finite-energy sum rules with a family of H\"older's inequalities, applied to an integrated hadronic spectral function~\cite{Li:2024frm}.
We present constraints on $a_\mu^{\mathrm{HVP,LO}}$ in Eq.~\eqref{eq:a_mu_QCD-result}, incorporating light-quark ($u,d,s$) contributions up to five-loop order in perturbation theory within the chiral limit, at LO in light-quark mass corrections, NLO in dimension-four QCD condensates, and at LO in dimension-six QCD condensates. 
Additionally, our results can be supplemented with contributions from charmonium and bottomonium states (\ie Eq.~\eqref{eq:inclusive_bound}) to provide QCD bounds that can be compared to those from other methodologies.
Our results help bridge the gap between lattice QCD and data-driven approaches, offering a path toward resolving the current tension in future investigations of the muon's anomalous magnetic moment.

\section*{Acknowledgments}
TGS is grateful for research funding from the Natural Sciences and Engineering Research Council of Canada (NSERC).

\vfill\eject
%%%%%%%%%%%%%%%

\end{document}